\documentclass{elsart}

\usepackage{natbib}

\usepackage{amssymb}

\begin{document}

\begin{frontmatter}



\title{Neutron star oscillations and QPOs during magnetar flares}


\author{Anna L. Watts}

\address{Max Planck Institut f\"ur Astrophysik,
  Karl-Schwarzschild-Str. 1, 85741 Garching, Germany}

\ead{anna@mpa-garching.mpg.de}
\ead[url]{http://www.mpa-garching.mpg.de/$\sim$anna}

 \& \author{Tod E. Strohmayer}
\address{Astrophysics Science Division, NASA Goddard Space
  Flight Center, Mail Code 662, Greenbelt, MD 20771, USA}

\begin{abstract}

The high frequency oscillations discovered in the tails of giant
flares from two magnetars are thought to be the first direct
detections of seismic vibrations from neutron stars.  The possibility
of starquakes associated with the giant flares triggering global vibrations
opens up the prospect of using seismology to study the
interior structure and composition of neutron stars.  This is a major
breakthrough in the study of the nature of matter under conditions of
extreme pressure.  In this paper we provide an up to date summary of the
observations and the theoretical framework, including a brief discussion of
gravitational wave searches for the QPOs.  We summarize the
status of alternative non-seismic mechanisms, and give a
critique of a
recent paper by Levin that argued against seismic vibrations as a
viable mechanism.    We conclude with an overview of current results using
the seismological technique that constrain
parameters such as the equation of state and crust structure.  

\end{abstract}


\end{frontmatter}


\section{Introduction}
\label{intro}

The Soft Gamma Repeaters are isolated compact objects that exhibit
regular gamma-ray flaring activity.  They are thought to be magnetars,
neutron stars with external magnetic fields in excess of $10^{14}$ G
and internal fields that could be as high as $10^{16}$ G
\citep{dun92, tho95}.  Decay of the field powers both
regular small-scale flares and rare giant flares that are orders of magnitude more
energetic.  Three such events have been observed in the era of
satellite-borne high energy detectors:  in 1979 from SGR 0526-66
\citep{maz79}, in 1998 from SGR 1900+14 \citep{hur99, fer99}, and in 2004 from SGR
1806-20 \citep{ter05, pal05, hur05}. The giant flares consist of a short hard flash (lasting less than a second),
following by a softer decaying tail that persists for several hundred
seconds.  Pulsations with periods of several seconds become
visible during the tail, revealing the spin period of the
neutron star.  They are due to a fireball of ejected plasma, trapped
near the stellar surface by the strong magnetic field \citep{tho95}.  

The giant flares are powered by catastrophic global reconfigurations
of the magnetic field.  However, the field is anchored to the charged
particles in the neutron star crust.  It had therefore long been
suspected that giant flares might trigger starquakes \citep{flo77,
  tho95, tho01, sch05} that would be sufficiently energetic to excite
global seismic vibrations \citep{dun98}.  The precise relationship
between  crust fracture and  flare is not yet clear.   In the model of
\citet{tho95} the field twists and puts the crust under strain, but
crustal rigidity prevents movement and reconnection.  Only when the
crust reaches its breaking strain is a flare triggered.  In the
alternative model of \citet{jon03} the crust deforms plastically as
the field twists, and flares occur when the field reaches an
instability point \citep{lyu03}.   Even if the
crust does not undergo brittle fracture, however, colossal and rapid
reconfiguration of the field alone could be enough to drive global
magneto-elastic vibrations.  

Neutron stars can sustain many types of oscillation restored by
various different 
forces.  Early calculations for neutron star models with a fluid
core and a solid crust indicated that the modes most likely to be excited
by a magnetar crustquake were the torsional shear oscillations of the crust,
with a fundamental frequency at $\approx 30$ Hz \citep{sch83, mcd88,
  str91a, dun98}.  These oscillations 
are primarily horizontal (as opposed 
to radial) and are restored by the shear modulus in the crust.    In what follows we use the quantum numbers $l$ and $m$ as
the standard labels for angular harmonics, and the quantum number $n$ to
denote the number of nodes in the radial eigenfunction that describes
the horizontal motion.  The precise harmonics excited would depend 
on fracture properties (location, geometry, speed), and subsequent
coupling and damping processes.  The means by which modes modulate the
x-ray lightcurve will also be important:  some modes may be excited
but not detectable due to the way that they couple to the external
field.  

\section{Observations}
\label{obs}

\subsection{Oscillations in the decaying tail}
\label{tail}

On December 27th 2004, the most energetic giant flare ever recorded
was detected from SGR 1806-20 \citep{ter05, pal05}.  Analysing data
of the decaying tail from the {\it Rossi X-ray Timing Explorer}
(RXTE) Proportional Counter Array (PCA), \citet{isr05} found a highly significant Quasi-Periodic Oscillation (QPO)
at 92 Hz.  The QPO was strongly rotational phase dependent,
appearing for only part of the rotational cycle away from the main
peak of the rotational pulse.  Weaker features at 18 and 30 Hz were
also found.  The 30 Hz and 92 Hz QPOs were in line with the
predictions of models for the fundamental torsional shear mode ($n=0, l=2$)
and one of the $n=0$ angular overtones.  The 18 Hz QPO was rather too low in
frequency to fit torsional shear mode models:  \citet{isr05} suggested
that it might instead by some kind of internal mode restored by the
strong field.  

Prompted by this discovery, \citet{str05} analysed RXTE PCA data from the
August 27th 1998 giant flare from SGR 1900+14.  Although the data are
not as high quality as for the SGR 1806-20 giant flare (there are data
gaps due to the satellite configuration), we found a strong transient QPO at
84 Hz.  Focusing on the rotational phase where the 84 Hz QPO was
strongest, and folding multiple cycles from the whole tail of the
flare, we found additional QPOs at 28, 54 and 155 Hz.  The set of four
QPOs could be fitted with a sequence of $n=0$ torsional shear modes of
different angular harmonic number $l$, with the 28 Hz QPO as the
fundamental.  

So by September 2005 QPOs with
similar frequencies and properties had been discovered in the decaying
tails of giant 
flares from two different objects.  The QPOs appeared to be transient,
but because it was necessary to fold multiple cycles to detect most of
the QPOs it was not clear when the oscillations were excited.  However
they were only seen clearly once the rotational pulsations were
established, suggesting an association with the surface.  In addition
both the 92 Hz QPO for SGR 1806-20 and all of the SGR 1900+14 QPOs
were strongly rotational phase dependent.  This raised interesting
questions.  Was the transience and rotational phase dependence caused
by material obscuring our view of the star (such as the rotating
fireball)?   And how did magnetic field geometry affect mode
patterns, survival of modes and their visibility?  

For the SGR 1900+14 flare the RXTE PCA data are the
only high time resolution data available for the tail.   For the SGR
1806-20 flare, several other high time resolution instruments might
have obtained useful data.  The SWIFT
Burst Alert Telescope (BAT) did record high time resolution data of the
initial part of the flare, but unfortunately the time resolution
dropped 20s into the flare, before the QPOs established themselves 
\citep{pal05}.  The {\it International Gamma-Ray Astrophysics Laboratory}
(INTEGRAL) also caught the flare.   However the time resolution of the
lightcurve obtained using the anti-coincidence shield on the
spectrometer was, at 50ms \citep{mer05}, too low for our purposes,  and
the countrate on 
the higher time resolution Joint European X-ray Monitor (JEM-X) was too low to be useful
\citep{bra05}.  

The only other spacecraft to get high time resolution data of the tail
of the flare was the {\it Ramaty High Energy Solar Spectroscopic
  Imager} (RHESSI), which was fortuitously pointing almost directly at
the flare.  RHESSI is a segmented detector covering both the
RXTE energy band and higher energies.  When both front and rear
segments are taken into account RHESSI recorded a higher
countrate than RXTE.  However, the rear segments were strongly
affected by albedo flux due to reflection from the Earth, which
complicated timing analysis. As such it was not possible to include events from
the rear segments when searching for QPOs above 50 Hz.   

Using the RHESSI data, \citet{wat06a} were able to confirm the RXTE detection of
the 92 Hz QPO.  We also found QPOs at 18 Hz and 26 Hz, at the same
rotational phase as the 92 Hz QPO.  A weak feature at 
30 Hz was not significant at the $3\sigma$ level once number
of trials were taken into account, so we were not able to make an
independent confirmation of this result (but see
below).   The most exciting discovery, however, was at higher
frequencies. Using photons with nominal energies in the range 100--200
keV (higher than the RXTE PCA energy band), we found a QPO 
at 625 Hz.   The QPO seemed to emerge
earlier in the tail than the 92 Hz QPO, and was strongest at a
different rotational phase.   A signal at 625 Hz was particularly
exciting as it is very close to the frequency predicted for the
$n=1$ overtones of 
the torsional shear modes \citep{pir05}. 

Once the RXTE data of the SGR 1806-20 giant flare went public we re-analysed this dataset to
check the RHESSI results and perform a thorough rotational phase
dependent search for additional QPOs \citep{str06}.  We were able to confirm the
RHESSI detections of the 18 and 26 Hz QPOs, and their rotational phase
dependence.  We were also able to confirm the presence of the 30 Hz
QPO found by \citet{isr05} and showed that it too is rotational
phase dependent.  A comparison of the spectra of photons detected by
RHESSI and RXTE indicated why the 30 Hz feature was weaker in the
RHESSI data:  the feature is strongest at low energies, a range in
which RHESSI is very noisy due (most likely) to scattering. 

Rotational phase dependent analysis also revealed new surprises in the
RXTE data.  The first was a QPO at 150 Hz, very close to one of the
frequencies detected in the SGR 1900+14 flare.  There was also an
additional QPO at 1840 Hz, close to the frequency expected for the
$n=3$ overtone of the torsional shear modes.  Most exciting, however, was a QPO at 625 Hz, the
frequency identified in the RHESSI dataset.  Compared to the RHESSI QPO, the RXTE QPO had lower
fractional amplitude, lower coherence and a different rotational
phase dependence.  It also appeared later in the tail of the flare, and in a
lower energy band.  In the context of the seismic mode model this
suggests two possibilities.  The first
is that we are seeing evolution of one mode due to changing emission
conditions.  At early times, the 
signal emits at high fractional amplitude in an energy band above
that of the RXTE PCA, so is only seen in RHESSI.  At later times, the
amplitude decays but so does photon energy, at which point the signal
is too weak to be seen in RHESSI (which recorded a lower countrate since we can only
use front segments for high frequency signals) but is detected in the
RXTE data.
A second possibility is that we are seeing two different $n=1$ modes,
since the different angular harmonics of the $n=1$ modes have very
similar frequencies \citep{pir05}.   

The 92.5 Hz QPO also showed strong evidence for variability in
amplitude and frequency over the course of the tail.  Whether this is
a property of the underlying oscillation, or indicative 
of variations in the emission mechanism and stellar environment, is as
yet unclear. Without better data it could be hard to pin this down.  

\subsection{Impulsive phase}

In addition to the variability observed in the tails of giant flares,
there is evidence for variability in the earlier, impulsive phase of
the flare. \citet{bar83}, analysing Prognoz and Venera data from the
first 200ms of the 1979 giant flare (the only high time resolution
data available), found evidence for variability at a
frequency of $\approx 43$ Hz. Geotail observations of the first 500ms of the SGR 1806-20 flare
indicate periodicity at $\approx 50$ Hz, but no similar phenomena are
seen in the SGR 1900+14 flare \citep{ter06}.  

Analysing variability in the impulsive phase is challenging, as there are
complicated dead-time effects.  However, the evidence is now mounting
for variability in this early phase, preceding the QPOs that appear
later in the tail once the surface of the star is visible.  Whether
the two phenomena are related is an interesting question.  The
timescales are similar, and it is possible that we are seeing the
event that triggers global vibrations or an earlier manifestation of
those oscillations.  As in the previous section this question will only be settled with better
data.  

\section{Global seismic vibration models}

\subsection{State of theoretical modelling}

\begin{table}
\center
\begin{tabular}{|c|c|l|}
\hline
\hline
SGR 1806-20 & SGR 1900+14 & Torsional shear mode identification\\
\hline
\hline
18$^*$ & & \\
\hline
26$^*$ & & \\
\hline
30$^*$ & 28 &  $n=0$, $l=2$\\
\hline
  & 53  &  $n=0$, $l=4$ \\
\hline
92$^*$  & 84 & $n=0$, $l=6$ \\
\hline
150 & & $n=0$, $l=10$ \\
\hline
 & 155 & $n=0$, $l=11$ \\
\hline
625$^*$  & & $n=1$ \\
\hline
1840 & & $n=3$ \\
\hline
\hline
\end{tabular}
\label{datasum}
\caption{Summary of QPO frequencies (in Hz) detected in the tails of the SGR 1806-20 and
  SGR 1900+14 giant flares.   QPOs with asterisks have been
  detected in both the RXTE and RHESSI datasets.  The 150 Hz and 1840
  Hz QPOs in the SGR 1806-20 flare have fractional amplitudes too low
  to be detected in RHESSI.  The third column indicates the
  identification as torsional shear modes on the basis of models that
  do not include crust-core coupling.  The mode identification that we have given for the $n=0$
  modes is consistent with that given in \citet{sam06}.  This differs
  very slightly from the identification given in our previous papers
  and in \citet{pir05}, and reflects the use of a (more accurate)
  spherical geometry in the more recent paper.}
\end{table}   

The frequencies of the QPOs are in good agreement, for the most part,
with models of torsional shear modes of neutron star crusts - in
accordance with the early theoretical suggestions.  The models used to
make these identifications were based on early calculations that
assumed free slip of the solid crust over the fluid neutron
star core \citep{han80, mcd88, str91a, str91b}. The models have since been improved
 to include the following effects:  gravitational
redshift; the effective boost to the shear modulus due to isotropic
magnetic pressure \citep{dun98}; up to date models of crust
composition \citep{hae94, pir05}; General Relativity and elasticity
\citep{sam06}.   Using
these free slip models, in which the crust moves independently of the
core,  one arrives at the mode identifications
summarized in Table \ref{datasum}.  Whilst most of the modes fit the
model well, the two lowest frequency modes in the SGR 1806-20 dataset
do not fit comfortably.  

The models described above have limitations in their treatment of the
magnetic field.  The assumption of an isotropic magnetic
field is clearly inadequate - we know that magnetars have strong
dipole field components and are also likely to have strong internal
toroidal fields \citep{bra06}.  The field also couples the crust to the
fluid core, modifying the boundary conditions at the interface.  A
number of papers have started to tackle these issues \citep{car86,
  mes01, gla06, lee06, sot06a, sot06b}.  Ultimately what one needs to do is
to compute global 
magneto-elastic modes of the star, taking full account of the
coupling.  An initial attempt to do this for a uniform field, using
simple slab geometry, was made by \citet{gla06}.  Slab geometry is
often used when computing crust modes, since curvature can be
neglected in the thin crust limit.  Clearly it has its limitations
when computing global modes, as it does not properly treat the conditions
in the centre of the star.  However, the results from this simple
model were extremely interesting.  Inclusion of the coupling resulted
in a set of modes whose amplitudes were strong in the crust and weak
in the core, with frequencies almost identical to the frequencies
computed when coupling was neglected.  However, the model also
generated additional  modes with lower
frequencies that could explain the 18 and 26 Hz QPOs detected in the
SGR 1806-20 flare.  

At this point it is appropriate to discuss the paper by
\citet{lev06} that argued against global seismic vibrations as a
viable mechanism for the magnetar QPOs.  The first part of the paper points out that if there is strong
displacement at the base of the crust then any movement of the crust
would necessarily cause Alfv\'en waves to propagate into the core.
This point is perfectly valid - one must, as we pointed out in the
previous paragraph, consider crust/core coupling. Levin then moves on
to consider 
the oscillations in the fluid core, using 
a toy model of a slab of fluid threaded by a position-dependent field.  The
particular model chosen possesses a continuous spectrum of apparently
singular modes.  Levin then adds a solid oscillating plate (the
`crust') to the top
of the slab, and argues that this will couple to the continuous
spectrum, causing energy to drain from the oscillating plate on a
rapid timescale.  On this basis he concludes that coupled oscillations
are also not viable as they will die away too rapidly.

The first problem with the toy model employed by Levin is that the continuous
spectrum is an artifact of the simple geometry, in particular the fact
that the model is unbounded in one 
dimension (the `z' dimension).  The model also neglects the fact that
there should be a `crust' on the lower edge of the slab as well.  Once
the proper boundaries are included one can compute perfectly viable
discrete modes:  this is demonstrated by the
computations in \citet{gla06}.  The varying-field problem, when
properly posed, may admittedly still
give rise to 
continuous spectra for
certain frequency bands.  However, the assertion made by Levin that this
automatically leads to rapid energy loss (based on an analogy with a
quantum mechanical problem) is certainly not proven.  Studies of
continuous spectra in differential rotation problems, for example,
show that the temporal behaviour of the collective physical
perturbation resulting from a continuous spectrum is complicated.
Decay of perturbations can be
polynomial - much slower than exponential decay - and discrete stable modes
can persist perfectly well even in the frequency bands occupied by the
continuous spectra \citep{wat03, wat04}. Levin's conclusions about the
viability of global seismic modes, based on the model presented in his
paper, are not sustainable.  

Accurate dissipation timescales for such global modes remain to be
calculated in detail.  However, estimates by \citet{sch83} and \citet{mcd88}
suggest that damping due to gravitational wave emission and neutrino
losses should be low.  \citet{bla89} estimated damping due to
Alfv\'en wave losses for crust-only modes and found that damping
drops off substantially for frequencies below several kHz.  This would
prolong global mode lifetimes in the frequency range of interest, a point also noted by \citet{dun98}.  As
mode calculations become more sophisticated all of these effects will
have to be revisited.  Other factors that need to be assessed include
dissipation within the core, viscous effects at the 
base of the crust, and the presence of crustal inhomogeneities. The
effect of the unusual magnetospheric conditions also
needs to be taken into account, since the presence of the trapped
fireball, which has a major impact on emission properties, will also
affect energy loss.   For the moment, however, the estimates are such
that the modes could survive long enough to explain the observations.   

The precise nature of behaviour at the crust/core interface also needs
more detailed consideration.  Boundary layer physics will be crucial
\citep{kin03}, as will the elastic properties at the base of the crust
\citep{pet98}.  Coupling will also depend rather sensitively on field
geometry:  the modes that survive may be those for which coupling to
the core is minimal.  Indeed the modes found by \citet{gla06} had very
weak amplitudes in the core.  The presence of a strong toroidal field in the
core would also increase its rigidity, rendering it less prone to
excitation.  Until all of these issues are addressed, the global
seismic vibration model remains the most promising mechanism for
explaining the QPOs, particularly given the difficulties faced by the
alternative mechanisms, discussed in Section \ref{alt}.

\subsection{Neutron star seismology}

Assuming that the global seismic vibration model is the correct
mechanism underpinning the oscillations, one can start to deduce the
properties of the magnetars' interiors.  This section summarises some
of the early findings. 

The frequencies of the fundamental $n=0$ torsional shear modes deduced
for SGRs 1806-20 and 1900+14 are different, the latter having a lower
fundamental.  This most probably reflects differences in the mass,
radius or magnetic field strength of the two stars (mass and radius
having a strong effect due to gravitational redshift). \citet{str05}
analysed the differences in 
mass and magnetic field that would be necessary to explain the different
frequencies, for various different equations of state \citep{lat01}.
Estimates of the magnetic field derived from timing \citep{woo02}
constrain the parameter space to rule out both very hard and very soft
equations of state.  For intermediate equations of state reasonable
results were obtained that support the finding that SGR 1806-20 has
a stronger field than SGR 1900+14.  

The possible identification of an $n=1$ radial overtone is
particularly exciting, as it enables us to estimate crust thickness.
This is on its own an independent constraint on the nuclear equation
of state.  In the Newtonian perturbation calculations of
\citet{han80}, one can show that in the thin crust limit the ratio of first radial overtone to 
fundamental frequency is proportional to the ratio of crust thickness to
stellar radius.   Subsequent Newtonian
perturbation calculations
included a correction to the frequency for gravitational redshift; a
factor which cancels when one takes the ratio of two frequencies.
\citet{str06} used this fact to compute mode frequencies for realistic
(non-thin) 
neutron star crust models in spherical geometry.  Using the SGR
1806-20 observations, 
assuming a fundamental at 30 Hz and a first overtone at 625 Hz, this
resulted in a crust thickness estimate of 0.1 --0.13 times the stellar
radius. Recent more detailed modelling is now
refining this initial value.  General relativistic mode
calculations by \citet{sam06} and \citet{sot06a}, which take into
account not only 
gravitational redshift but also the effect on crust thickness
associated with relativistic models, have already resulted in revised
estimates.  Efforts to include crust/core coupling will doubtless
lead to further changes \citep{gla06, sot06b}.  This is nonetheless the
first time that 
there has been the
possibility of making a direct measurement of crust thickness in a
neutron star.  

This is particularly important for strange star models.  Strange stars
are hypothetical compact objects, proposed by \citet{wit84}, composed
entirely of stable strange quark matter.   Their predicted properties
are very similar to those of neutron stars.  However, models predict
that their crusts should differ, the major difference being that they
are much thinner than neutron star crusts \citep{alc86, jai06}.
\citet{wat06b} have recently computed torsional shear mode frequencies
for strange star crust models to see whether they are compatible with
the observations.  The frequencies found differ substantially from the
frequencies found for neutron star crusts, and there is a particular
problem with the overtone frequencies.  As
expected, the thin crust, coupled with effects due to the strong field, pushes the
frequencies of the overtones up.  Even given very poor constraints on
some of the strange matter parameters, the overtone frequencies are
far too high to explain a QPO at 625 Hz.  These findings need to be
verified using more sophisticated models, but this may well be a
robust method to rule out strange stars.

\section{Gravitational wave searches}

Non-axisymmetric oscillations of neutron stars will emit
gravitational radiation. Magnetar starquakes are therefore of interest
as potential sources for detectors such as the {\it Laser
  Interferometer Gravitational-Wave Observatory} (LIGO).  Estimates of
the gravitational wave emission
of crustal torsional shear vibrations  indicate that
the signals would be far too weak to be detected even by Advanced LIGO
\citep{sch83}.  However, if the oscillations also involve the core,
prospects for detection improve.  At the time of the 2004 giant flare
from SGR 1806-20, only the 4 km interferometer at LIGO's Hanford site
was taking data. LIGO scientists are at present analysing this data to
place upper limits on any periodic gravitational wave signal at the
detected frequencies.  Details of the analysis were presented by
\citet{mat06}, and upper limits should be forthcoming in the near
future.  Since the 2004 event, the sensitivity of LIGO has improved still
further, and (particularly with all the detectors operating) prospects
for setting even more stringent upper limits on future events are
good.

\section{Alternatives to seismic models}
\label{alt}

\subsection{Modes of a debris disk}

In a recent paper, \citet{wan06} reported the discovery of a debris
disk around a magnetar, the
Anomalous X-ray Pulsar 4U 0142+61.  Many neutron stars in binary
systems exhibit high frequency QPOs that are thought to originate in the
accretion disk \citep{van06}.  It has therefore been suggested that a
similar phenomenon might be responsible for the magnetar QPOs.  

It is important to note that there is no evidence as yet for
fallback/debris disks around SGRs, although giant flares clearly expel
plasma.  There is also no evidence from the kHz QPOs of normal neutron
stars for the  
sequence of frequencies that we find in the SGRs, or for any
rotational phase dependence.  However, most importantly there are
major differences between the accretion 
disks in neutron star binaries and the debris disk found by
\citet{wan06}.  The critical property for timing purposes is the inner
disk radius.  For the neutron stars that show high frequency QPOs the
inner disk radius is close the neutron star, and the orbital frequencies
in the disk can exceed 1000 Hz.  All current kHz QPO models rely on
the presence of rapid orbital frequencies within the disk
\citep{van06}.  The inner 
disk radius for the debris disk around the magnetar is, by comparison,
several solar radii.  There are no rapid orbital frequencies in this
system.    This alone seems to rule out disk models as a viable mechanism
for the magnetar QPOs.  

\subsection{Modes of the magnetosphere}

Vibrations within the magnetosphere have been suggested by several
authors as a possible alternative to global seismic vibrations of the
star.  A simple estimate of Alfv\'en speeds in the magnetosphere
suggests that the frequencies of global Alfv\'en modes would be too
high. However, there may be 
alternative magnetospheric mechanisms.  A recent paper by
\citet{bel06} on magnetar coronae indicates that there should be very
high frequency  quasi-periodic oscillations ($\sim 10$ kHz) in the
inner coronal electric current due to the corona being in a state of
self-organized criticality. Lower frequency oscillations would require
larger lengthscales and so would probably have to involve the outer
corona: but emission is much weaker in this region, making it
difficult to achieve the measured QPO amplitudes.  However, further
work on the state of the corona in the immediate aftermath of a
giant flare is required before a coronal mechanism can be ruled out
conclusively.  

\section{Conclusions}

Neutron stars are the best laboratories for extreme physics in the
Universe.  Understanding the nature of matter in the cores of neutron
stars would be an immense step forward in nuclear physics, and the
magnetar QPO detections represent the first serious 
opportunity to study this region using seismology.   Early analysis has already shown the
immense potential of this technique to constrain the nuclear equation of state
and crust thickness of neutron stars, findings that may rule out strange star
models.  Whilst there are many theoretical issues still to be
resolved, the most urgent requirement is to improve our observational
capability so as to obtain the best possible data of any future giant
flare. The potential scientific payoff easily justifies the effort.

\section{Acknowledgments}

We are grateful to Luca Matone for discussions about the gravitational
wave search, Andrei Beloborodov for his thoughts on magnetospheric
oscillation mechanisms, and Deepto Chakrabarty for pointing out that a
large inner disk radius would preclude a disk as a mechanism for the
QPOs.  ALW is also grateful to COSPAR for a travel grant that enabled
her to attend the Beijing meeting.


\end{document}